\documentclass[fleqn,usenatbib,useAMS]{mnras}

\usepackage{graphicx}	% Including figure files
\usepackage{amsmath}	% Advanced maths commands
\usepackage{amssymb}	% Extra maths symbols
\usepackage{multicol}        % Multi-column entries in tables
\usepackage{bm}		% Bold maths symbols, including upright Greek
\usepackage{pdflscape}	% Landscape pages

\newcommand\hii{H\,{\sc ii} \,}
\newcommand\degree{^{\circ}}
\usepackage{cjhebrew}
\newcommand{\mpo}[1]{\textcolor{blue}{#1}} %%%%%%%%%%%%%%%%%%%%%%%%%%%%%%%%%%%%%%%%%%%%%%%%%% 

\usepackage[T1]{fontenc}
\usepackage{ae,aecompl}

\usepackage{newtxtext,newtxmath}
\date{Last updated 2020 June 10; in original form 2013 September 5}

\pubyear{2022}

\title[Rectangular core-collapse supernova remnants]{Rectangular 
core-collapse supernova remnants: application to Puppis A \\ }

\author[D. M.-A.~Meyer et al.]
       {
       D. M.-A.~Meyer\thanks{E-mail: dmameyer.astro@gmail.com}$^{1}$,  
       P.~F.~Velázquez$^{2}$, O.~Petruk$^{3,4}$, 
       A.~Chiotellis$^{5}$, M.~Pohl$^{1,6}$, A.~Camps-Fariña$^{2,7}$, \newauthor
       M.~Petrov$^{8}$, E.~M.~Reynoso$^{9}$, J.~C.~Toledo-Roy$^{2}$, E.~M.~Schneiter$^{10}$,
       A.~Castellanos-Ramírez$^{11}$,
       \newauthor A. Esquivel$^{2}$\\ 
       $^{1}$ Universit\" at Potsdam, Institut f\" ur Physik und Astronomie, 
                 Karl-Liebknecht-Strasse 24/25, 14476 Potsdam, Germany \\
       $^{2}$ Instituto de Ciencias Nucleares, Universidad Nacional Aut\'onoma 
                 de M\'exico, CP 04510, Mexico City, Mexico \\ 
       $^{3}$ Institute for Applied Problems in Mechanics and Mathematics, NAS of Ukraine, 
                 Naukova 3-b, 79060 Lviv, Ukraine  \\      
       $^{4}$ Astronomical Observatory of Taras Shevchenko National University of Kyiv, 3 Observatorna str. Kyiv, 04053, Ukraine \\
       $^{5}$ Institute for Astronomy, Astrophysics, Space Applications and Remote Sensing, 
                 National Observatory of Athens, 15236, Penteli, Greece  \\                        
       $^{6}$ DESY Platanenallee 6, 15738 Zeuthen, Germany \\  
       $^{7}$ Departamento de F\'{i}sica de la Tierra y Astrof\'{i}sica, Universidad Complutense de Madrid, 28040 Madrid, Spain \\
       $^{8}$ Max Planck Computing and Data Facility (MPCDF), Gießenbachstrasse 2, D-85748 Garching, Germany \\
       $^{9}$ Instituto de Astronom\'{i}a y F\'{i}sica del Espacio (IAFE), Av. Int. G\"uiraldes 2620, 
               Pabellón IAFE, Ciudad Universitaria, \\1428, Buenos Aires, Argentina \\
        $^{10}$ Departamento de Materiales y Tecnolog\'{i}a, FCEFyN-UNC, Av. V\'elez Sarsfield 1611, C\'ordoba, Argentina \\
        $^{11}$ Universidad Nacional Aut\'onoma de M\'exico, Instituto de Astronom\'{i}a, Ap. 70-264, CDMX, 04510, M\'exico.
       }

\begin{document}
\label{firstpage}
\pagerange{\pageref{firstpage}--\pageref{lastpage}}
\maketitle

% Abstract of the paper
\begin{abstract}
Core-collapse supernova remnants are the gaseous nebulae of galactic
interstellar media (ISM) formed after the explosive death of massive stars. 
Their morphology and emission properties depend both 
on the surrounding circumstellar structure shaped by the stellar wind-ISM interaction
of the progenitor star and on the local conditions of the ambient medium. 
In the warm phase of the Galactic plane ($n\approx 1\, \rm cm^{-3}$, $T\approx 8000\, \rm K$), an 
organised magnetic field of strength $7\, \mu \rm G$ has profound consequences on the morphology 
of the wind bubble of massive stars at rest. In this paper we show through 2.5D magneto-hydrodynamical 
simulations, in the context of a Wolf-Rayet-evolving $35\, \rm M_{\odot}$ star, that it affects 
the development of its supernova remnant. 
When the supernova remnant reaches its middle age ($15$$-$$20\, \rm kyr$), it adopts a tubular 
shape that results from the interaction between the isotropic supernova ejecta and the anisotropic, 
magnetised, shocked stellar progenitor bubble into which the supernova blast wave expands. 
Our calculations for non-thermal emission, i.e. radio synchrotron and inverse Compton radiation, reveal that such supernova remnants can, due to projection effects, appear as rectangular objects in certain cases. 
This mechanism for shaping a supernova remnant is similar to the bipolar and elliptical planetary nebula production by wind-wind interaction in the low-mass regime of stellar evolution. If such a rectangular core-collapse supernova remnant is created, the progenitor star must not have been a runaway star.
We propose that such a mechanism is at work in the shaping of the asymmetric 
core-collapse supernova remnant Puppis~A. 
\end{abstract}

% Select between one and six entries from the list of approved keywords.
% Don't make up new ones.

%%%%%%%%%%%%%%%%%%%%%%%%%%%%%%%%%%%%%%%%%%%%%%%%%%%%%%%%%%%%%%%%%%%%%%%%%%%%%%%%%%%%%%%%%%%
%%%%%%%%%%%%%%%%%%%%%%%%%%%%%%%%%%%%%%%%%%%%%%%%%%%%%%%%%%%%%%%%%%%%%%%%%%%%%%%%%%%%%%%%%%%
%%%%%%%%%%%%%%%%%%%%%%%%%%%%%%%%%%%%%%%%%%%%%%%%%%%%%%%%%%%%%%%%%%%%%%%%%%%%%%%%%%%%%%%%%%%

\section{Introduction}
\label{sect:intro}

Core-collapse supernova remnants are the result of
the explosive death of massive stars ($\ge 8\, \rm M_{\odot}$). 
These astrophysical events are rare. However, the mass and
energy release in supernova explosions triggers a feedback process of 
substantial importance in the behavior and evolution of the interstellar medium 
(ISM). This feedback enriches the ISM with heavy elements 
and drives turbulence, regulating future star formation processes ~\citep{hopkins_mnras_417_2011}.
Core-collapse supernova remnants result from the propagation
of the energetic blastwave produced by supernova explosions into the 
surroundings of a high-mass star, which prior to the explosion  
shaped its environment as structured, concentric layers of hot and cold 
gas, either from the stellar wind or from the ISM, separated by shocks~\citep{woosley_araa_24_1986,smartt_araa_47_2009}. 
Hence, the morphology of supernova remnants is a function of both the density 
distribution of the circumstellar medium (partly reflecting the past evolution 
of the progenitor star) and the local condition of the ISM. 
Therefore, studying the morphology of supernova remnants is key to 
understand stellar evolution, to constrain local properties of the ISM and to
examine the physics of particle acceleration at shocks~\citep{langer_araa_50_2012}. 
The classic picture for core-collapse supernova remnants can be found 
in~\citet{chevalier_apj_258_1982, chevalier_apj_344_1989,truelove_apjs_120_1999}.

The vast variety of morphologies found in 
core-collapse supernova remnants is a mystery. Most of them substantially  
deviate from the spherically-symmetric solution~\citep{aschenbach_aa_341_1999,arias_aa_622_2019,arias_aa_627_2019,
domcek_aa_659_2022,zhou_2022arXiv220313111Z}. 
Many reasons can be invoked to explain this observational trend, like 
asymmetric stellar winds, or the presence of inhomogeneities in the 
surrounding medium. Both cases can induce a non-spherical expansion of 
the supernova shock wave.  
Indeed, the life span of the progenitor star, the successive stellar 
evolutionary phases, and the properties of the corresponding stellar winds
depend on the star's initial mass~\citep{maeder_araa_38_2000}. The stellar
rotation~\citep{szecsi_aa_658_2022} and the existence of one or more 
companions~\citep{sana_sci_337_2012} are additional factors that strongly 
modify the structure of the stellar wind bubble and, consequently, that of the 
supernova remnant. 
The density distribution of the local, primeval ISM, such as the presence 
of a dense molecular component close to the site of the explosion, is another 
source of anisotropy prone to affect the propagation of the supernova 
blastwave~\citep{fesen_mnras_481_2018,boumis_mnras_512_2022}.

Over the past decades, the wind bubble problem has spurred the interest of the 
numerical astrophysical community~\citep{garciasegura_1996_aa_316ff,garciasegura_1996_aa_305f,  
freyer_apj_594_2003,dwarkadas_apj_630_2005,freyer_apj_638_2006,eldridge_mnras_367_2006}. 
Moreover,  
the large morphological variety of supernova remnants provided a natural continuation of that computational  effort, both in the high-mass~\citep{dwarkadas_apj_667_2007,vanveelen_aa_50_2009} and 
low-mass~\citep{chiotellis_aa_537_2012,chiotellis_mnras_435_2013,broersen_mnras_441_2014,chiotellis_galax_8_2020,chiotellis_mnras_502_2021} regime of the progenitors. 
Many processes have been identified as sources of asymmetry in the 
circumstellar medium of massive stars, and consequently in their subsequent 
supernova remnants. 
%s
Amongst others, the role of radiation in shaping the surroundings of 
massive stars is investigated in~\citet{toala_apj_737_2011,geen_mnras_448_2015}. 
The runaway motion of a fraction of massive 
stars~\citep{blaauw_bain_15_1961,gies_apjs_64_1987,Moffat1998} 
distorts wind bubbles into bow shocks~\citep{gull_apj_230_1979,wilkin_459_apj_1996,
scherer_mnras_493_2020,baalmann_aa_634_2020,baalmann_aa_650_2021,baalmann_paper_2022,herbst_ssrv_218_2022}, 
naturally providing an asymmetric environment to the expanding shock wave and 
generating asymmetric supernova remnants~\citep{franco_pasp_103_1991,rozyczka_mnras_261_1993,
brighenti_mnras_273_1995, 
brighenti_mnras_277_1995,fang_mnras_464_2017,zhang_apj_867_2018,fang_mnras_435_2013,
yang_raa_20_2020,meyer_mnras_496_2020,meyer_mnras_507_2021}, themselves able to govern  
the morphology of pulsar wind nebulae~\citep{2022arXiv220603916M}. 
Recently, \citet{2022arXiv220303369D} noted a softening of the non-thermal particle spectra when the shock wave goes 
through shocked circumstellar material. 
Similarly, the role of material expelled throughout pulsating mass-loss events 
occurring during the last century of the progenitor's life has been 
highlighted in the context of the asymmetries in the expanding shock wave 
of Cassiopea~A~\citep{orlando_aa_645_2021,2022arXiv220201643O}. 
More generally, the lengthening effects of an organised magnetic field on the morphology 
of stellar wind bubble was analysed by~\citet{vanmarle_584_aa_2015}.

\begin{table*}
	\centering
	\caption{
	List of models in this study. All simulations assume a non-rotating static $35\, \rm M_{\odot}$ star
	at solar metallicity. The table indicates
	for each simulation their label, the number of grid
	points used, and the strength of the ISM magnetic field (in $\mu \rm G$). 
	}
	\begin{tabular}{lccr}
	\hline
	${\rm {Model}}$         &  Grid mesh              &   $B_{\rm ISM}$ ($\mu\, \rm  G$)  
                         	&    Description     \\ 
	\hline   
	Run-35-HD-0-CSM         &  $2500 \times 5000$     &  0  
	&  Circumstellar medium in unmagnetised ISM         \\
	Run-35-MHD-0-CSM        &  $2500 \times 5000$     &  7  
	&  Circumstellar medium in magnetized ISM           \\
	Run-35-HD-0-SNR         &  $3000 \times 6000$     &  0  
	&  Supernova remnant in unmagnetised ISM            \\
	Run-35-MHD-0-SNR        &  $3000 \times 6000$     &  7  
	&  Supernova remnant in magnetized ISM              \\	
	\hline    
%	\hline 
	\end{tabular}
\label{tab:table1}
\end{table*}

This study is a first step in tackling the problem of the particular appearance 
of the core-collapse supernova remnant Puppis~A, which exhibits a rhomboid 
shape, inside of which several filaments are arranged 
orthogonal to each other, 
providing the overall appearance of nested squares and rectangles. 
This specific arrangement is observable at many wavebands, spanning from 
the infrared~\citep{arendt_apj_725_2010} and the optical/X-rays regimes~\citep{hwang_apj_635_2005} 
to non-thermal radio emission~\citep{hewitt_apj_759_2012}. 
%\
We employ magneto-hydrodynamical (MHD) simulations to explore the single 
O-type progenitor scenario for Puppis~A proposed in~\citet{reynoso_mnras_464_2017}. 
In our chosen sequence of events, the supernova explosion takes place 
within a circumstellar medium that has been self-consistently pre-shaped 
by the previous release of mass and energy along the successive 
stellar evolutionary phases (main-sequence, red supergiant and Wolf-Rayet) into the ISM. 

Our work is organised as follows. In Section~\ref{sect:method}, we review the 
numerical methods and the initial conditions employed to perform numerical
magneto-hydrodynamical simulations for the circumstellar medium and subsequent 
supernova remnant of a massive progenitor at rest in a magnetised ISM. 
The asymmetries in the supernova remnant are directly obtained as a result of 
the interaction of the supernova shock wave with the pre-shaped magnetised 
circumstellar medium of the progenitor star~\citep{vanmarle_584_aa_2015}. 
The evolution of the remnant is presented in Section~\ref{sect:results}. 
We investigate the non-thermal and thermal appearance of our supernova remnant 
in Section~\ref{sect:discussion}, and further discuss our findings in the 
context of the core-collapse supernova remnant Puppis~A. 
Finally, we summarize our conclusions in Section~\ref{sect:conclusion}.

%%%%%%%%%%%%%%%%%%%%%%%%%%%%%%%%%%%%%%%%%%%%%%%%%%%%%%%%%%%%%%%%%%%%%%%%%%%%%%%%%%%%%%%%%%%
%%%%%%%%%%%%%%%%%%%%%%%%%%%%%%%%%%%%%%%%%%%%%%%%%%%%%%%%%%%%%%%%%%%%%%%%%%%%%%%%%%%%%%%%%%%
%%%%%%%%%%%%%%%%%%%%%%%%%%%%%%%%%%%%%%%%%%%%%%%%%%%%%%%%%%%%%%%%%%%%%%%%%%%%%%%%%%%%%%%%%%%

\section{Method}
\label{sect:method}

In this section we provide the reader with a description of the numerical methods utilized 
to perform simulations of asymmetric core-collapse supernova remnants of rectangular appearance. 
We also briefly summarize the radiative transfer recipes for non-thermal emission used in the 
analysis of our simulated models.

\subsection{Circumstellar medium}
\label{sect:method_csm}

First, the circumstellar medium of a massive star is simulated. The 2.5D models 
are carried out in cylindrical coordinates ($R$,$z$,$\theta$) with a static grid, 
under the simplifying assumption of $z$-axis symmetry.  
The dimensions of the grid is $[O,R_{\rm max}]\times[z_{\rm min},z_{\rm max}] 
\times[0,2\pi]$, with a mesh that is uniform along the $R$- and $z$-directions. 
It is constructed of $N_{\rm R}=2500$ and $N_{\rm z}=5000$ grid zones and it extends 
to $R_{\rm max}=z_{\rm max}=150\, \rm pc$ and $z_{\rm min}=-150\, \rm pc$. The uniform 
resolution of the mesh is therefore $\Delta=|z_{\rm max}-z_{\rm min}|/N_{\rm z} 
=R_{\rm max}/N_{\rm R}$. 
Reflective conditions are assigned along the $z$-axis (i.e. at the boundary $R=0$) 
and outflow conditions at the other boundaries, respectively. 
The computational domain is initially filled with gas having the properties of the 
warm phase of the ISM in the Galactic plane of the Milky Way, with number density 
$n_{\rm ISM}\approx 0.79\, \rm cm^{-3}$ and $T_{\rm ISM}\approx 8000\, \rm K$. 
The magnetization of the ISM is imposed by the geometry of the domain with magnetic field parallel 
to the symmetry axis $Oz$.

\textcolor{black}{
Our choice of an organised, parallel magnetic field is justified, since even though 
the typical description of the ISM is that of a turbulent 
plasma~\citep{passot_apj_455_1995,zank_apj_887_2019}, 
a large scale \mpo{coherent} magnetic field is present in the spiral disc 
galaxies \citep[see models and observations 
of][]{yusefzadeth_apj_320_1987,baryshnikova_aa_177_1987,brown_apj_663_2007,beck_aa_470_2007,2012ApJ...757...14J}. 
This large-scale field is responsible for, e.g. the alignment of supernova remnants 
with the enrolled arms of the Milky Way~\citep{gaensler_apj_493_1998} and affects 
the development of stellar wind bubbles of massive stars~\citep{vanmarle_584_aa_2015}. 
The turbulent component of the magnetic field has a significant strength overall \citep{2012ApJ...761L..11J}, but it includes fluctuations with wavelength up to about $100$~pc with an energy density that falls off with wavenumber, $k$. Radiation modeling of young supernova remnants suggests that the coherence length of the magnetic field can indeed be large compare to the size scale of our simulations \citep{2020MNRAS.496.2448P}. Then the random magnetic field on small scales acts as an additional 
isotropic, negligible pressure to the ambient medium in which circumstellar shocks propagate.  
}

The stellar wind is imposed in a sphere of radius $r_{\rm in}=20$ grid zones, following 
the method of~\citet{comeron_aa_338_1998}~\citep[see also][]{mackey_apjlett_751_2012,meyer_2014bb}. 
The stellar wind density is imposed assuming the profile
\begin{equation}
	\rho_{w} = \frac{ \dot{M} }{ 4\pi r^{2} v_{\rm w} },
\label{eq:wind}
\end{equation}
where $r=\sqrt{R^2+z^2}$ is the radial distance from the pre-supernova location, $\dot{M}$ the mass-loss rate and $v_{\rm w}$ 
the wind terminal velocity. The mass-loss rate $\dot{M}$ is time-dependently interpolated from 
the stellar evolutionary track of~\citet{ekstroem_aa_537_2012}, the wind velocity is obtained 
from the escape velocity using the law of~\citet{eldridge_mnras_367_2006}. 
This star successfully evolves through a long hot main-sequence phase, a short cold red 
supergiant phase and finishes its life via a hot Wolf-Rayet phase. The main properties of 
the star are presented in great details in~\citet{meyer_mnras_502_2021}. 
The evolution of the wind-ISM interaction is followed throughout the entire star life, from 
the onset of the main-sequence phase at $t=0$ to the pre-supernova time at $t=5.41\, \rm Myr$.

We perform two simulation models of similar ISM and stellar wind properties, the difference 
being the strength of the ISM magnetic field, taken to be $B_{\rm ISM}=0\, \mu \rm G$ 
(hydrodynamical model) and $B_{\rm ISM}=7\, \mu \rm G$ (magneto-hydrodynamical model), 
see Table~\ref{tab:table1}.

\subsection{Supernova remnant}
\label{sect:method_snr}

Once we know the structure of the circumstellar medium at the pre-supernova time, 
we calculate the initial ejecta-wind interaction in 
spherical symmetry in the radial range [$O$;$r_{\rm out}$]. 
The ejecta material fills the inner $r \le r_{\rm max} < r_{\rm out}$ 
region of the domain. The corresponding density field is defined as a constant density,  
\begin{equation}
   \rho_{\rm core} =  \frac{1}{ 4 \pi n } \frac{ \left(10 E_{\rm ej}^{n-5}\right)^{-3/2}
 }{  \left(3 M_{\rm ej}^{n-3}\right)^{-5/2}  } \frac{ 1}{t_{\rm max}^{3} },
   \label{sn:density_1}
\end{equation}
within $r \leq r_{\rm core}$, and a power-law density profile 
\begin{equation}
   \rho_{\rm max}(r) =  \frac{1}{ 4 \pi n } \frac{ \left(10 E_{\rm
ej}^{n-5}\right)^{(n-3)/2}  }{  \left(3 M_{\rm ej}^{n-3}\right)^{(n-5)/2}  } \frac{ 1}{t_{\rm max}^{3} } 
\bigg(\frac{r}{t_{\rm max}}\bigg)^{-n},
   \label{sn:density_2}
\end{equation}
in the $r_{\rm core} < r \le r_{\rm max}$ region, respectively, 
with $n=11$~\citep[see][]{chevalier_apj_258_1982,truelove_apjs_120_1999}.

The ejecta speed follows homologous expansion, $v(r) = r/t$, and 
\begin{equation}
   v_{\rm core}(r_{\rm core}) = \bigg(  \frac{ 10(n-5)E_{\rm ej} }{ 3(n-3)M_{\rm ej} } \bigg)^{1/2}.
   \label{sn:vcore}
\end{equation}
The forward shock of the supernova blastwave at $r_{\rm max}$ has a flow speed of 
$v_{\rm max}=30000\, \rm km\, \rm s^{-1}$, which constrains the time of the simulation 
start $t_{\rm max} = r_{\rm max}/v_{\rm max}$. The value of $r_\mathrm{max}$ is 
calculated using the method described in~\citet{vanveelen_aa_50_2009}. 
The value of $r_{\rm max}$ is arbitrary as long as the mass it embeds 
is smaller than that of the ejecta, and $t_{\rm max}$ is determined 
assuming $v_{\rm max}=30000\, \rm km\, \rm  s^{-1}$.
In Eq.~\ref{sn:density_1}-\ref{sn:vcore}, $E_{\rm ej}=5 \times 10^{50}\, \rm erg$ is the 
total energy released by the supernova explosion, and $M_{\rm ej}=10.12\, \rm M_{\odot}$ 
is the ejecta mass, calculated as the pre-supernova stellar mass minus a neutron star 
mass of $M_{\mathrm{NS}}=1.4\, \rm M_{\odot}$. 
Finally, the 1D ejecta-wind interaction is mapped onto a 2.5D computational domain 
before the forward shock of the blastwave runs into the termination shock of the 
stellar wind bubble, see method in~\citet{meyer_mnras_493_2020,meyer_mnras_502_2021}. 
We assume that supernova ejecta and stellar wind are not magnetised.

\subsection{Governing equations}
\label{sect:method_eq}

The time evolution of the gas is calculated within the frame of ideal 
magneto-hydrodynamics~\citep{meyer_mnras_496_2020}, described by, 
\begin{equation}
	   \frac{\partial \rho}{\partial t}  + 
	   \bmath{\nabla}  \cdot \big(\rho\bmath{v}) =   0,
\label{eq:mhdeq_1}
\end{equation}
\begin{equation}
	   \frac{\partial \bmath{m} }{\partial t}  + 
           \bmath{\nabla} \cdot \Big( \bmath{m} \textcolor{black}{\otimes} \bmath{v}  
           \textcolor{black}{-} \bmath{B} \textcolor{black}{\otimes} \bmath{B} + \bmath{\hat I}p_{\rm t} \Big)  \bmath{0},
\end{equation}
\begin{equation}
	  \frac{\partial E }{\partial t}   + 
	  \bmath{\nabla} \cdot \Big( (E+p_{\rm t})\bmath{v}-\bmath{B}(\bmath{v}\cdot\bmath{B}) \Big)  
	  = \Phi(T,\rho),
\label{eq:mhdeq_3}
\end{equation}
and
\begin{equation}
	  \frac{\partial \bmath{B} }{\partial t}   + 
	  \bmath{\nabla} \cdot \Big( \bmath{v}  \textcolor{black}{\otimes} \bmath{B} - \bmath{B} \textcolor{black}{\otimes} \bmath{v} \Big)  =
	  \bmath{0},
\label{eq:mhdeq_4}
\end{equation}
which stand for the mass continuity, momentum conservation, energy conservation equations and for 
the evolution of the magnetic field, respectively. 
In the above relations, $\bmath{m}=\rho\bmath{v}$ is the vector momentum, $\rho$ the gas density, 
$v$ the gas velocity, $p_{t}$ the total pressure and, 
\begin{equation}
	E = \frac{p}{(\gamma - 1)} + \frac{ \bmath{m} \cdot \bmath{m} }{2\rho} 
	    + \frac{ \bmath{B} \cdot \bmath{B} }{2},
\label{eq:energy}
\end{equation}
the total energy of the gas. 
The definition of the sound speed $c_{\rm s} = \sqrt{ \gamma p/\rho }$ closes the system 
of equations, with $\gamma=5/3$ the adiabatic index. Optically-thin radiative cooling and 
photo-heating $\Phi(T,\rho)$ are included into the equations via the prescriptions 
of~\citet{meyer_2014bb}, where $T$ is the gas temperature.

The equations are solved using the {\sc pluto} 
code\footnote{http://plutocode.ph.unito.it/}~\citep{mignone_apj_170_2007,
migmone_apjs_198_2012,vaidya_apj_865_2018}. This study uses the Godunov-type solver 
utilised in~\citet{meyer_mnras_506_2021}, made of the shock-capturing HLL Rieman 
solver~\citep{hll_ref} and the eight-wave finite-volume algorithm~\citep{Powell1997} 
which ensures that the magnetic field remains divergence-free. 
The numerical method is made of a third-order Runge-Kutta time integrator, used 
together with the minmod flux limiter and the WENO3 interpolation scheme. 
The time-step is controlled by the Courant-Friedrich-Levy, initialised to 
$C_{\rm cfl}=0.1$~\citep{meyer_mnras_507_2021}.

\begin{figure*}
        \centering
        \includegraphics[width=0.9\textwidth]{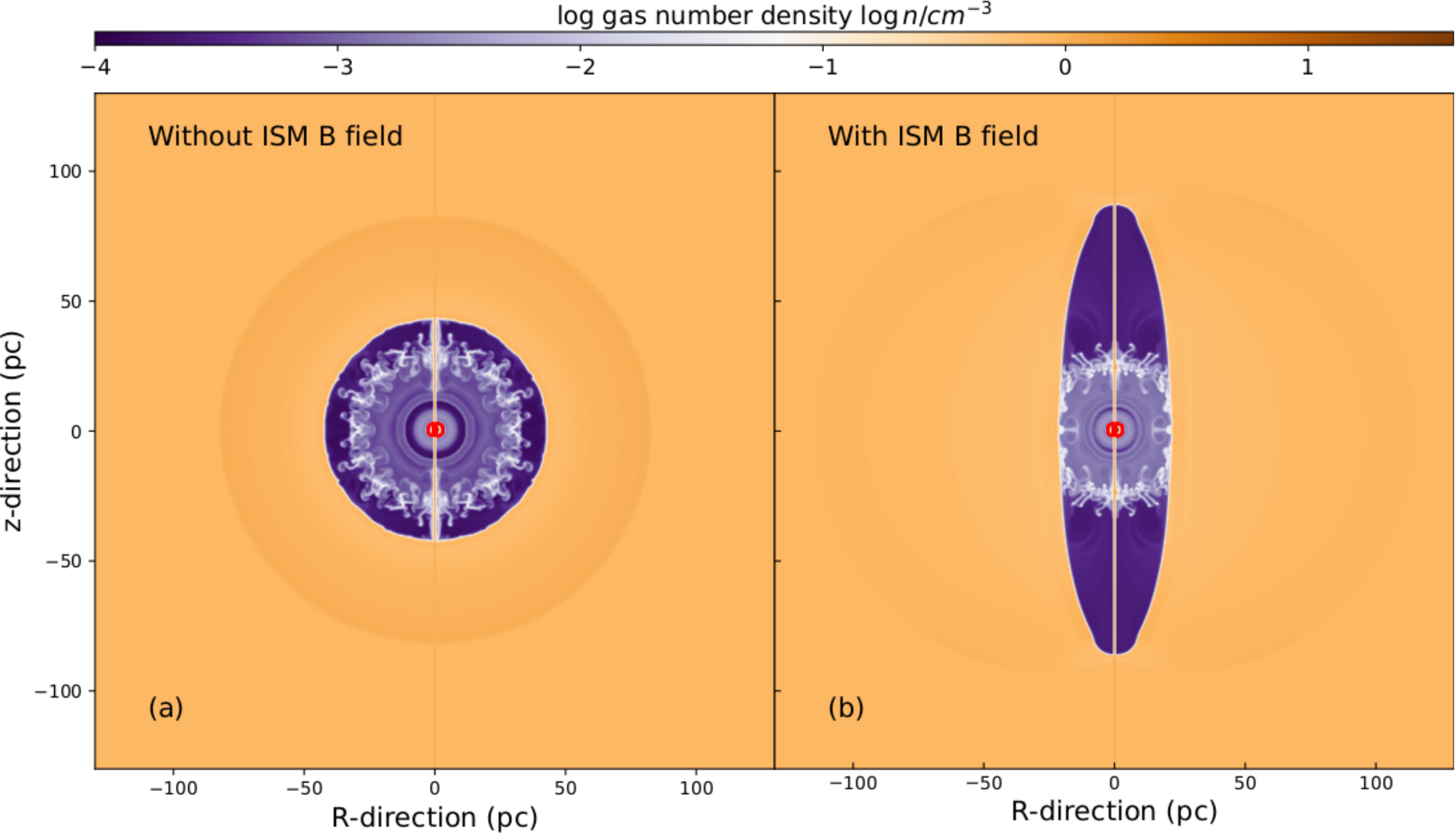}  \\        
        \caption{
        Number density field (in $\rm cm^{-3}$) in our models for the circumstellar 
        medium of a static $35\, \rm M_{\odot}$ star, shown at the supernova time, 
        without (a) and with (b) ISM magnetic field. 
        The red contour marks the region of the circumstellar medium where is 
        supernova material and energy is injected. 
        }
        \label{fig:plot_csm}  
\end{figure*}

\subsection{Non-thermal emission maps}
\label{sect:method_rt}

For further comparison between our MHD models and available 
observational data, we perform radiative transfer calculations of selected time instances of the 
supernova remnant evolution, for non-thermal emission such as synchrotron and inverse Compton 
emission. 
Regarding the synchrotron radio emission, we use the method detailed in~\citet{meyer_mnras_502_2021}. 
It first assumes a spectrum of the non-thermal electrons present in the post-shock region of the 
propagating supernova blastwave,  
\begin{equation}
        N(E) = K E^{-s},
        \label{eq:N}  
\end{equation}
where $s=2$, $K\propto n$ with $n$ the gas number density, and with $E$ the electron energy. The emission coefficient 
at frequency $\nu$ then reads 
\begin{equation}
        j_{\rm sync}(\nu) \propto n B_{\perp}^{ (s+1)/2 } \nu^{ -(s-1)/2 },
        \label{eq:coeff}  
\end{equation}
where $B_{\perp}$ stands for the magnetic
field component perpendicular to the observer's line-of-sight, $l$. The viewing angle $\theta_{\rm obs}$ 
is the inclination angle of the supernova remnant with respect to the plane of the sky. 
The emission maps are calculated using the radiative transfer code 
{\sc RADMC-3D}\footnote{https://www.ita.uni-heidelberg.de/$\sim$dullemond/software/radmc-3d/} 
which permits ray-tracing integration of arbitrary emission coefficients, providing non-thermal emission maps of
synchrotron intensity, 
\begin{equation}
        I = \int_{\rm SNR} j_{\rm sync}\left(\theta_{\rm obs}\right)  dl.
        \label{eq:intensity}  
\end{equation}

\begin{figure*}
        \centering
        \includegraphics[width=0.96\textwidth]{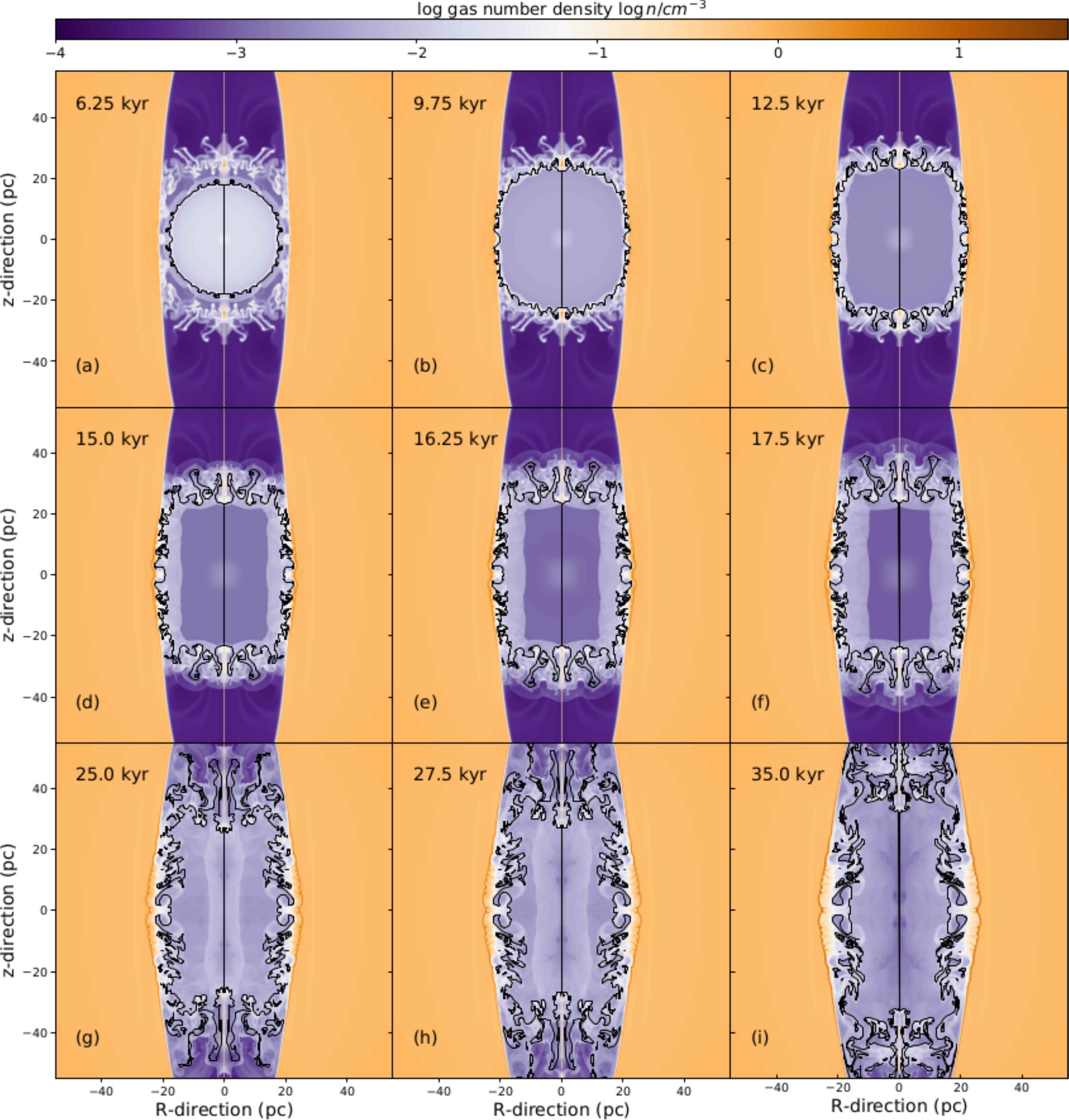}  \\        
        \caption{
        Number density field (in $\rm cm^{-3}$) in our models for the supernova remnant of 
        a static $35\, \rm M_{\odot}$ star in a magnetized ISM. 
        The black contour marks the region of the remnant 
        made of $50\%$ of ejecta in number density. 
        }
        \label{fig:plot_snr}  
\end{figure*}

\begin{figure*}
        \centering
        \includegraphics[width=0.95\textwidth]{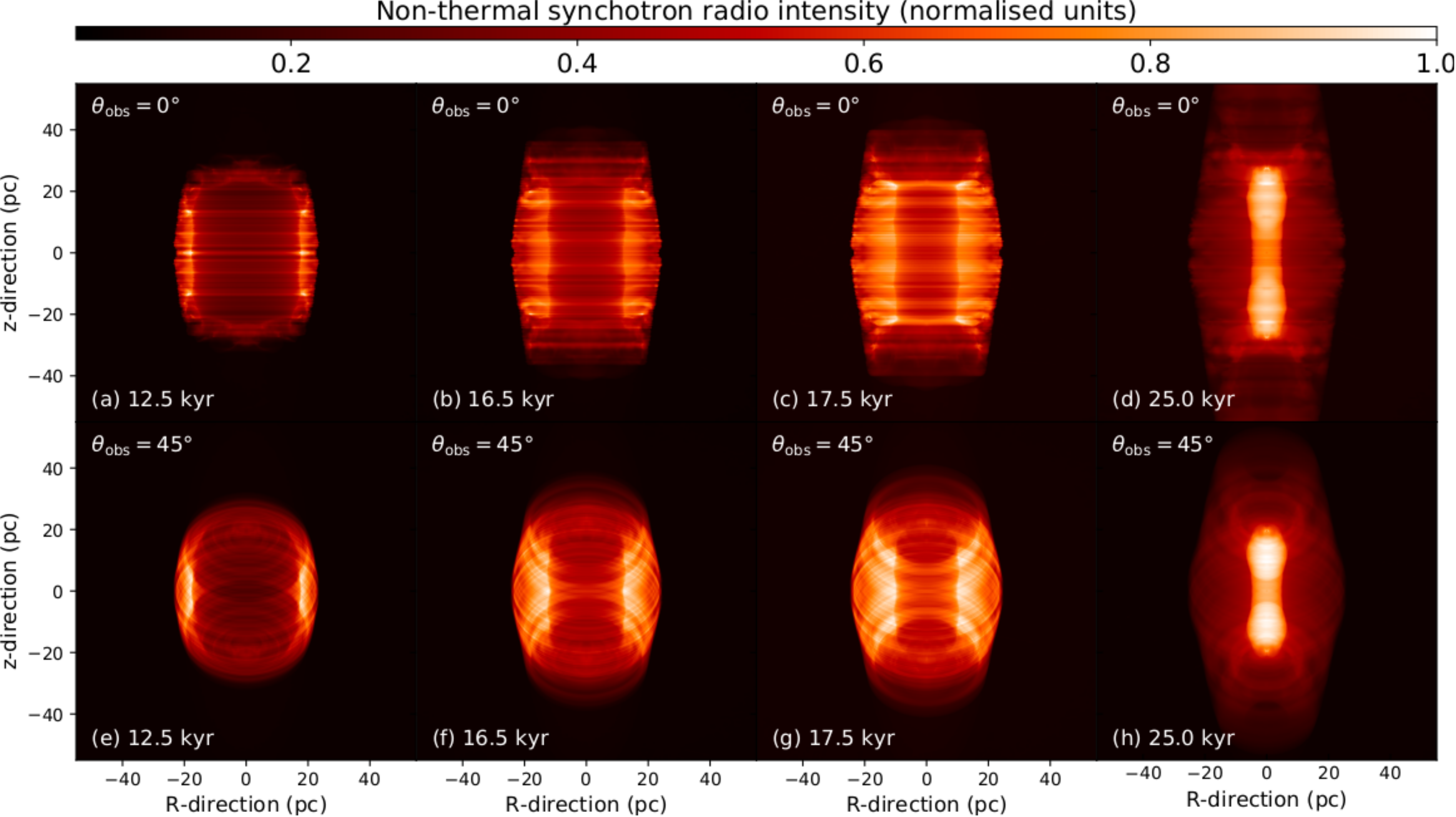}  \\        
        \caption{
        Normalized non-thermal radio synchrotron intensity maps of our supernova 
        remnant model of a static $35\, \rm M_{\odot}$ star, seen according 
        viewing angles of $\theta_{\rm obs}=0\degree$ (top panels) and 
        $\theta_{\rm obs}=45\degree$ (bottom panels). 
        The images are displayed for time instances $12.5\, \rm kyr$ 
        (left panels) to $25.0\, \rm kyr$ (right panels) after the 
        supernova explosion. The images are centered onto the location of 
        the supernova explosion.  
        }
        \label{fig:plot_sync}  
\end{figure*}

Similarly, we calculate emission maps for inverse Compton emission by assuming 
that the electron spectrum (Eq.~\ref{eq:N})
extends to a sharp cut off at a very high energy $E_\mathrm{max}$.
We make use of the emission coefficient at a given photon energy $\epsilon$, 
\begin{equation}
        j_{\rm IC} = \int_{0}^{E_{\rm max}} N(E) \Lambda(E) dE 
        \propto n \epsilon^{-(s-1)/2}, %\frac{ n }{ v^{-b} }, 
        \label{eq:emission_IC}  
\end{equation}
with $\Lambda(E)$ an integral that is a function of the target photon 
field assumed to be the cosmic microwave background, as established in~\citet{petruk_aa_499_2009}. 
This method has been shown to be accurate from the Thomson to the Klein-Nishina 
regimes. We further assume that the maximum accelerated energy, $E_{\rm max}$, 
is the same everywhere in the supernova remnant. 
The normalized emission maps are obtained by performing the integral,  
\begin{equation}
        I = \int_{\rm shocks} j_{\rm IC}  dl,
        \label{eq:intensity_IC}  
\end{equation}
for the shocked material in the vicinity of the forward and reverse shocks of the 
supernova remnant.

These simple, recipes permit the calculation of predictive non-thermal emission 
from supernova remnant simulations~\citep[e.g.][]{castellanos_mnras_508_2021}. 
A more careful determination of 
synchrotron and inverse Compton fluxes of our supernovae remnants models would 
require the explicit knowledge of the accelerated particle distribution as done 
in~\citet{brose_aa_593_2016,brose_aa_627_2019,sushch_apj_926_2022,2022arXiv220303369D}.

\subsection{Thermal emission maps}
\label{sect:method_th}

Last, maps of optical H$\alpha$ and soft X-rays emission from the core-collapse 
supernova remnants are calculated. The procedure is similar as for the non-thermal 
emission and uses different emission coefficients. 
The optical H$\alpha$ intensity reads, 
\begin{equation}
        I = \int_{\rm SNR} j_{\rm H\alpha}  dl,
        \label{eq:intensity_Ha}  
\end{equation}
with, 
\begin{equation}
        j_{\rm H\alpha} = \Big( 1.21 \times 10^{-22} T^{-0.9} \Big)  n_{\rm H}^{2}, 
        \label{eq:emission_Ha}  
\end{equation}
where $n_{\rm H}$ is the gas hydrogen number density~\citep{meyer_mnras_450_2015}. 
The soft X-ray emission is calculated as,  
\begin{equation}
        I = \int_{\rm SNR} j_{\rm XR}  dl,
        \label{eq:intensity_XR}  
\end{equation}
with, 
\begin{equation}
        j_{\rm XR} = \Lambda(T) n_{\rm H}^{2}, 
        \label{eq:emission_XR}  
\end{equation}
with $\Lambda(T)$ the $0.1-1.0\, \rm keV$ emissivity for diffuse ISM obtained with the 
{\sc xspec} software~\citep{arnaud_aspc_101_1996}.

%%%%%%%%%%%%%%%%%%%%%%%%%%%%%%%%%%%%%%%%%%%%%%%%%%%%%%%%%%%%%%%%%%%%%%%%%%%%%%%%%%%%%%%%%%%
%%%%%%%%%%%%%%%%%%%%%%%%%%%%%%%%%%%%%%%%%%%%%%%%%%%%%%%%%%%%%%%%%%%%%%%%%%%%%%%%%%%%%%%%%%%
%%%%%%%%%%%%%%%%%%%%%%%%%%%%%%%%%%%%%%%%%%%%%%%%%%%%%%%%%%%%%%%%%%%%%%%%%%%%%%%%%%%%%%%%%%%

\section{Results}
\label{sect:results}

In this section, we present our (non-)magnetized models for the pre-supernova circumstellar 
medium of static massive stars and for the evolution of their subsequent supernova 
remnant, together with predictions by radiative transfer calculation of their thermal and 
non-thermal appearance.

\subsection{Asymmetric circumstellar medium}
\label{sect:result_csm}

In Fig.~\ref{fig:plot_csm} we show the density field (in $\rm cm^{-3}$) in the 
simulations for the circumstellar medium of a static $35\, \rm M_{\odot}$ star 
without (panel a, Run-35-HD-0-CSM) and with an ISM magnetic field  
(panel b, Run-35-MHD-0-CSM), see Table~1 for the models properties. 
The hydrodynamical simulation (Fig.~\ref{fig:plot_csm}a) exhibits the typical 
morphology of a stellar wind bubble generated by a massive static star. 
From the star to the outermost part of the bubble, we have: (1) the freely-expanding 
Wolf-Rayet stellar wind, (2) the unstable shell of shocked Wolf-Rayet wind 
sweeping the hot diluted cavity of RSG material, (3) the contact discontinuity separating the main-sequence 
material and the ISM gas, (4) the large layer of cold, dense shocked ISM material, and (5) the forward shock sweeping the unperturbed ambient medium.
The overall structure is globally spherically-symmetric~\citep{weaver_apj_218_1977}, 
although the instabilities in the last Wolf-Rayet wind break this 
symmetry~\citep{garciasegura_1996_aa_305f,garciasegura_1996_aa_316ff}. 
We refer the reader interested in further details on the physics of stellar 
wind bubbles around massive stars to the literature devoted to this 
problem~\citep{freyer_apj_594_2003,dwarkadas_apj_630_2005,freyer_apj_638_2006}.

Fig.~\ref{fig:plot_csm}b displays the pre-supernova circumstellar medium of 
the massive star in a magnetised ISM. Only the strength of the magnetic field 
differs from Run-35-HD-0-CSM, in which it was set to $B_{\rm ISM}=0\, \mu\rm G$. 
The evolution of the contact discontinuity during the main-sequence phase of the 
stellar evolution is 
asymmetric and elongated as a consequence of being inhibited along the direction normal to the 
vertical ISM field lines~\citep{vanmarle_584_aa_2015}.
Note that the density at the forward shock of the wind bubble is smaller in the 
magnetized case, since the magnetic pressure reduces the compression ratios of shocks (Fig.~\ref{fig:plot_csm}a,b).
The distribution of the RSG stellar wind is strongly affected by the 
shape of the contact discontinuity and it adopts the same oblong configuration. 
Similarly, the expansion of the Wolf-Rayet stellar wind, initially spherical, is 
modified by the morphology of the distribution of the termination shock of 
the stellar wind bubble, see also~\citet{meyer_mnras_496_2020}.

\subsection{Asymmetric supernova remnant}
\label{sect:result_snr}

In Fig.~\ref{fig:plot_snr} we show as a time sequence the evolution of the  
supernova blastwave into the circumstellar medium of a $35\, \rm M_{\odot}$ star 
in a magnetized ISM (model Run-35-MHD-0-CSM). The figures display the number density field of the 
supernova remnant between $6.25\, \rm kyr$, (a), 
and $35.0\, \rm kyr$, (i), after the explosion. The thin black contour marks 
the location where $50\%$ of the number density is contributed by ejecta. 
The supernova shock wave first expands into the freely-expanding Wolf-Rayet 
wind and approaches the unstable shell of shocked Wolf-Rayet 
gas (Fig.~\ref{fig:plot_snr}a). The shell has begun to interact with the 
contact discontinuity of the main-sequence wind bubble and loses its spherical 
shape~\citep{vanmarle_584_aa_2015}. 
As the blastwave continues expanding, the shock wave 
loses sphericity (Fig.~\ref{fig:plot_snr}b) and adopts an 
elongated morphology mirroring that of the cavity of the stellar 
wind bubble previously shaped by the progenitor star (Fig.~\ref{fig:plot_snr}c). 
This is precisely the mechanism that our study aims at highlighting, within the 
context of the core-collapse supernova remnant Puppis A.

Partial reflection of the blastwave at the shell of Wolf-Rayet material sends gas
back towards the centre of the explosion. Shock reflection is at 
work at the walls of the stellar wind cavity and, to a lesser degree, 
at the basis of the cylinder of Wolf-Rayet wind, which still expand and 
interact with the unperturbed wind material (Fig.~\ref{fig:plot_snr}d). 
A similar mechanism is at play at a 
stellar wind bow shock~\citep{meyer_mnras_450_2015} or a cold ISM region~\citep{ferreira_478_aa_2008,castellanos_mnras_508_2021}.
The tubular morphology of the wind cavity consequently imposes the reflected 
shocks of supernova ejecta a cylindrical shape, which persists as it propagates 
towards the center of the supernova remnant (Fig.~\ref{fig:plot_snr}e). 
Simultaneously, the forward shock of the supernova is transmitted 
through the Wolf-Rayet shell keeps on expanding in the vertical direction 
of the cavity, as well as in the region of shocked ISM gas (Fig.~\ref{fig:plot_snr}f). 
After the vertically-reflected waves join near the axis of symmetry of the 
supernova remnant, the process of reflection and transmission of the shock 
front continues and the supernova ejecta, Wolf-Rayet, red supergiant and 
main-sequence materials mix within a thin unstable zone encompassing a 
dense central region of ejecta (Fig.~\ref{fig:plot_snr}g-i).

\begin{figure*}
        \centering
        \includegraphics[width=0.95\textwidth]{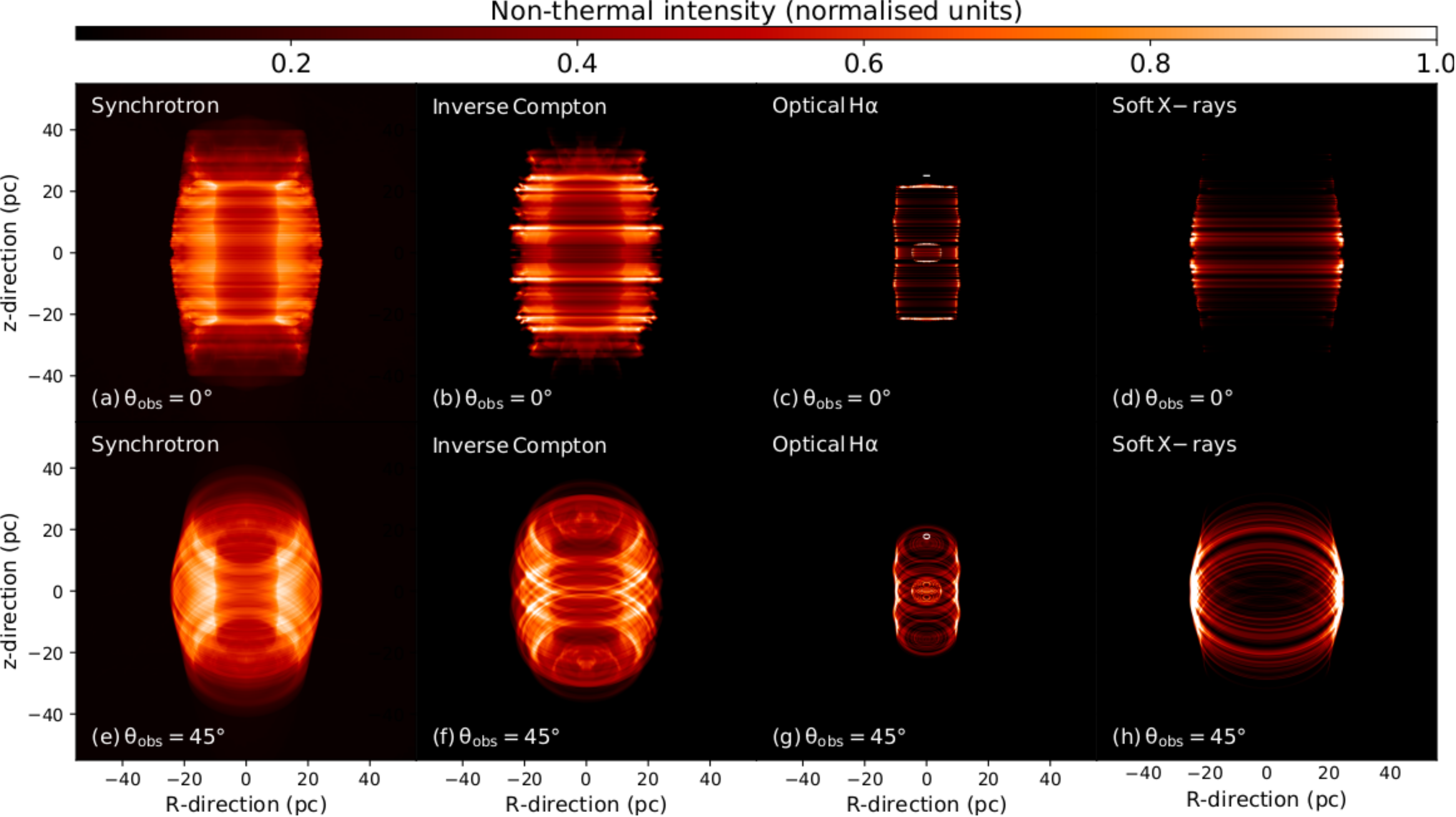}  \\        
        \caption{
        Normalized non-thermal radio synchrotron intensity (left)
        and inverse Compton (middle left)
        as well as thermal optical H$\alpha$ (middle right) and thermal X-rays
        (right) emission maps of our supernova
        remnant model of a static $35\, \rm M_{\odot}$ star in a magnetized 
        medium at time $17.5\, \rm kyr$ after the explosion, seen according 
        viewing angles of $\theta_{\rm obs}=0\degree$ (top) and 
        $\theta_{\rm obs}=45\degree$ (bottom). The images are center onto the location of 
        the supernova explosion. 
        }
        \label{fig:plot_IC}  
\end{figure*}

\subsection{Non-thermal emission}
\label{sect:diss_emission}

Fig.~\ref{fig:plot_sync} shows the normalized radio synchrotron maps for our simulation Run-35-MHD-0-CSM at the time $12.5\, \rm kyr$ 
(left panels), $16.5\, \rm kyr$ (left middle panels), $17.5\, \rm kyr$ (right middle panels), 
and $25.0\, \rm kyr$ (right panels) after the supernova explosion, respectively.
The images are obtained using the emission 
coefficient described in Section~\ref{sect:method} and assuming viewing angles of 
$\theta_{\rm obs}=0\degree$ (top panels) and $\theta_{\rm obs}=45\degree$ (bottom panels). 

When the shock wave hits the cavity walls, the radio appearance of the supernova remnant is that of two parallel bright bars aligned with the direction of the ISM magnetic 
field (Fig.~\ref{fig:plot_sync}a). This loss of sphericity of the expanding blastwave is also 
clearly visible with a viewing angle of $\theta_{\rm obs}=45\degree$ (Fig.~\ref{fig:plot_sync}e). 
As the blastwave propagates further and is reflected, we would see an outer hexagon produced by the ejecta penetrating 
the shock ISM material, inside of which two reflected parallel waves demarcate a bright 
bilateral region (Fig.~\ref{fig:plot_sync}b,f). 
When the supernova blastwave reflects onto the shell of Wolf-Rayet stellar wind, 
it converges toward to the centre of the explosion. The region of shocked 
ejecta consequently adopts the shape of a bright rectangle embedded in an hexagon, 
at least for a viewing angle slightly tilted
with respect to the plane of the sky 
(Fig.~\ref{fig:plot_sync}c,g). 
However, once the reflected waves join in the center of the supernova remnant, a very bright 
elongated region of shocked ejecta forms, and the remnant 
loses its rectangular shape (Fig.~\ref{fig:plot_sync}d,h).

Fig.~\ref{fig:plot_IC} compares synchrotron and inverse Compton emission for 
the viewing angles  $\theta_{\rm obs}=0\degree$ (top) and $\theta_{\rm obs}=45\degree$ (bottom)
(left columns, respectively) at an age of 17.5~kyr, when the rectangular 
synchrotron morphology is most pronounced. The ring-like features which appear for a viewing angle $\theta_{\rm obs}=45\degree$ 
are artifacts produced by the 2D nature of the simulations. 
The synchrotron map with $\theta_{\rm obs}=0\degree$ traces the compressed magnetic 
field in the reflected ejecta and in the ejecta penetrating the shocked ISM 
bubble (Fig.~\ref{fig:plot_IC}a). The inverse Compton emission is faint within 
the central rectangle. In contrast, the brightest inverse Compton emission originates from the dense material at the unstable ejecta/wind interface, perpendicular 
to the direction of the ISM magnetic field (Fig.~\ref{fig:plot_IC}b). 
For both non-thermal emission 
mechanisms the viewing angle is an important factor for the rectangular morphology (Fig.~\ref{fig:plot_IC}e,f). 

\begin{figure*}
        \centering
        \includegraphics[width=0.99\textwidth]{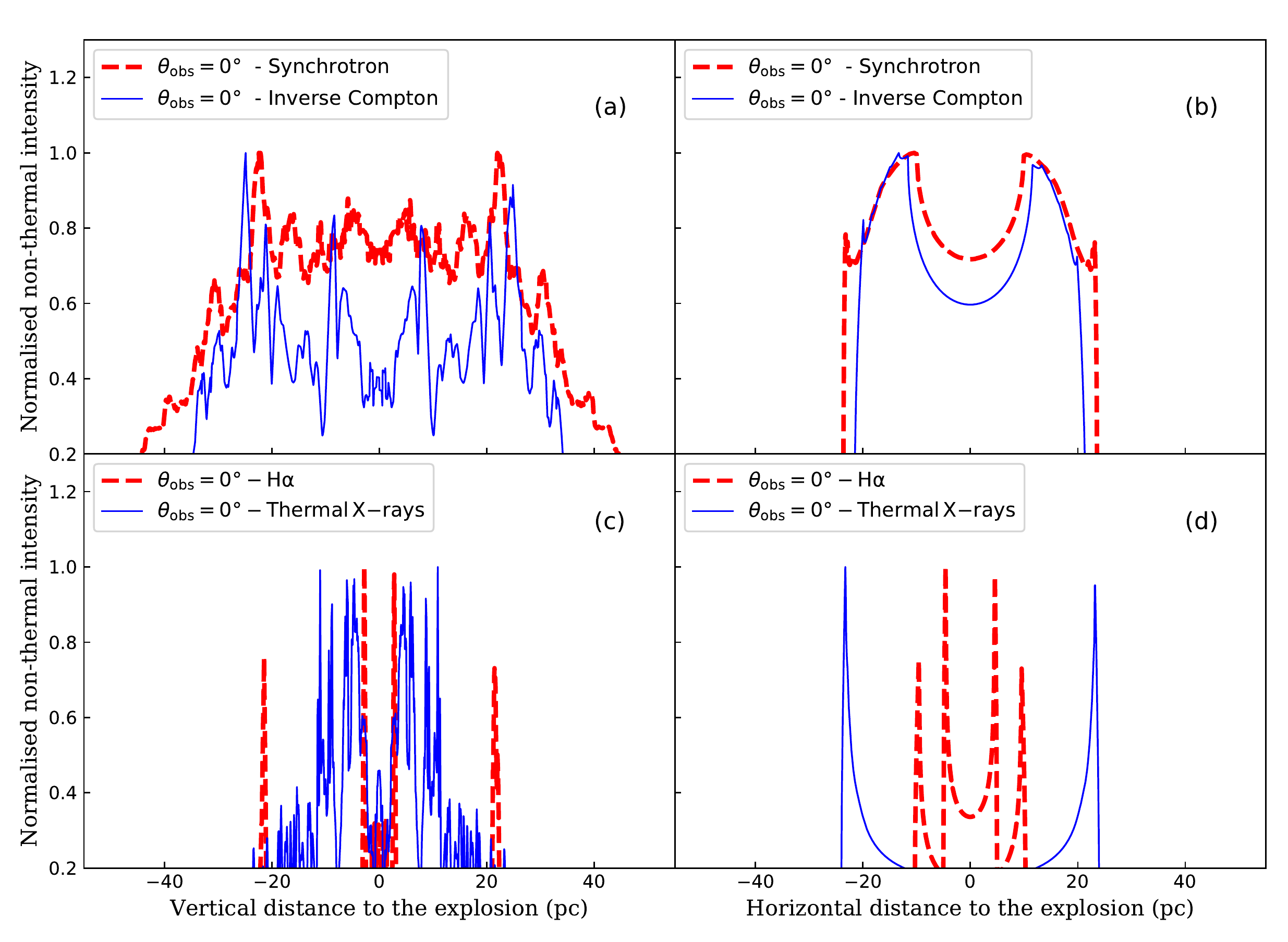}       
        \caption{
        Cuts through the normalized synthetic non-thermal (a,b) 
        and thermal (c,d) emission maps of our 
        rectangular supernova remnant at time $17.5\, \rm kyr$, 
        taken through the center of the explosion, along the vertical 
        direction (left column) and the horizontal direction (right column), 
        under a viewing angle of $\theta_{\rm obs}=0\degree$. 
        }
        \label{fig:emission_cuts}  
\end{figure*}

Fig.~\ref{fig:emission_cuts} displays a series of cross-sections taken 
vertically (left) and horizontally (right) through the intensity 
maps in Fig.~\ref{fig:plot_IC} for a viewing angle of $\theta_{\rm obs}=0\degree$. 
The top panels of the figures display the cuts for 
synchrotron (thick dashed red line) and inverse Compton (thin solid blue line) 
emission. 
Fig.~\ref{fig:emission_cuts_2} is as Fig.~\ref{fig:emission_cuts}, but assuming 
a viewing angle of $\theta_{\rm obs}=45\degree$. 
The vertical slices show that the synchrotron cavity, extending from $-20\, \rm pc$ 
to $20\, \rm pc$, is centre-filled by bright rings produced by the axisymmetric 
character of our MHD simulation. This is not seen  
in the inverse Compton intensity, regardless of the viewing angle $\theta_{\rm obs}$ 
(Fig.~\ref{fig:emission_cuts}a and Fig.~\ref{fig:emission_cuts_2}a). 
The horizontal cross-sections highlight the differential expansion of the supernova  
blastwave into the tubular magnetised circumstellar medium which induce the rectangular 
morphology, i.e. the brightest peaks of the horizontal slice sit about 
$20\, \rm pc$ from the center of the explosion while the vertical extent
reaches up to $ 40\, \rm pc$.

\subsection{Thermal emission}
\label{sect:therm_emission}

\begin{figure*}
        \centering
        \includegraphics[width=0.99\textwidth]{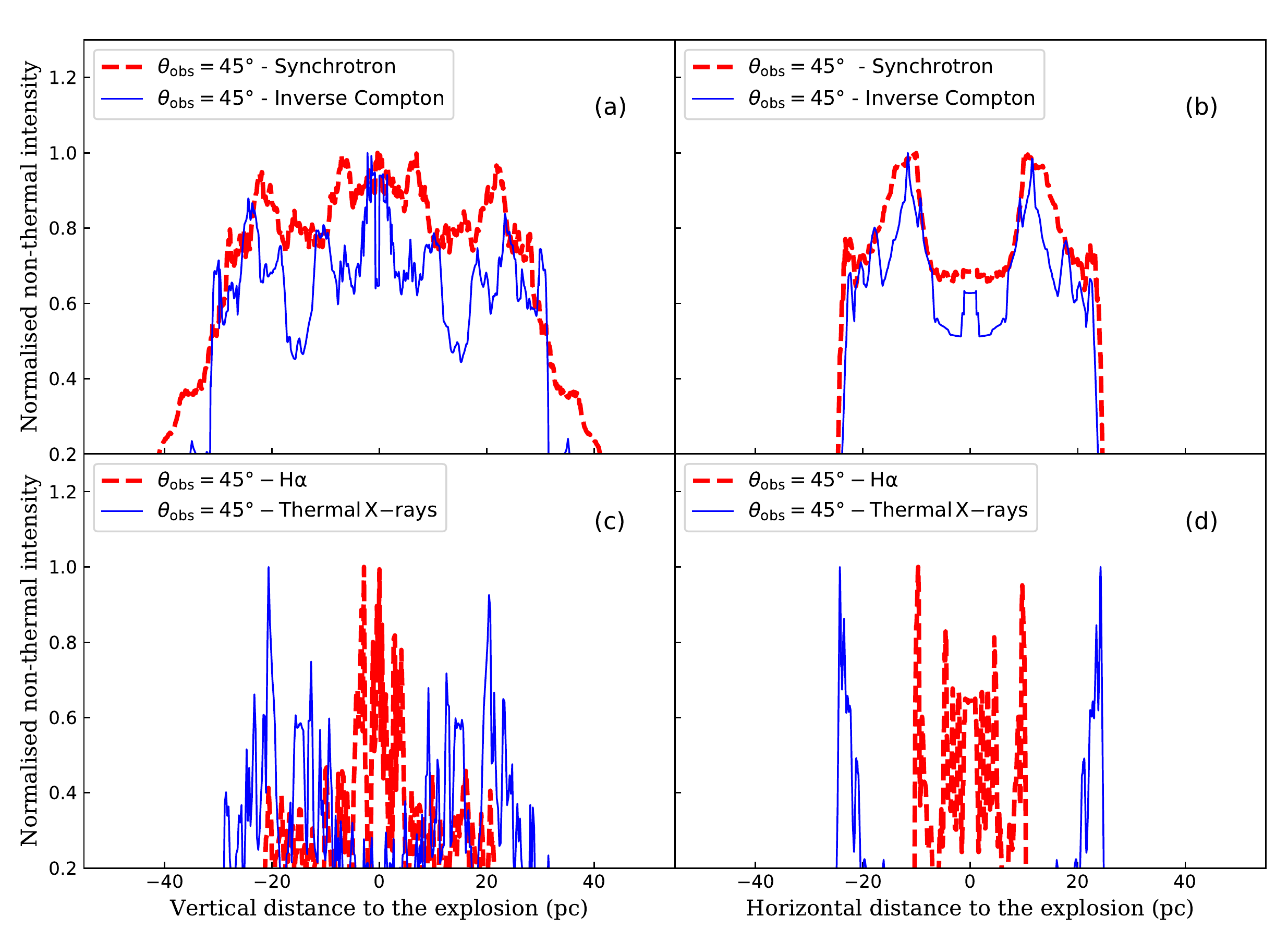}       
        \caption{
        As Fig.~\ref{fig:emission_cuts}, but with $\theta_{\rm obs}=45\degree$. 
        }
        \label{fig:emission_cuts_2}  
\end{figure*}

The right-hand series of panels in Fig.~\ref{fig:plot_IC} display optical H$\alpha$ (panels c,g)
and thermal X-ray (panels d,h) emission maps at the time of a prominent rectangular 
synchrotron morphology. 
The optical image traces the reverse shock of the blastwave being reverberated towards 
the center of the explosion, giving the remnant its peculiar morphology,  
with a central dense ring that reflects the initial inner ejecta profile. 
The transmitted shock passing through the magnetised and elongated walls of the stellar wind cavity is not visible
despite the high post-shock density, since the high temperature in this 
region forbids generous optical emission. 
The soft thermal X-ray emission basically originates from the supernova blastwave which 
penetrates into the walls of the cavity, where the gas is both dense and hot. 
The morphology of the projected emission is square, as is the overall shape of Puppis A. 
The bottom panels of Figs.~\ref{fig:emission_cuts} and~\ref{fig:emission_cuts_2}  
plot cross-sections taken vertically (panel c) and horizontally (panel d) through 
the optical (thick dashed red line) and thermal X-ray (thin solid blue line) 
emission maps, respectively. 
Both plots additionally highlight that the H$\alpha$ photons are mostly emitted in the region limited by
the reverberated forward shock of the supernova blastwave. 
Its vertical and horizontal extension away from the center of the explosion differs greatly 
so that in projection the region looks like a rectangle. 
Conversely, the thermal X-ray emission finds its origin in the supernova shock wave 
penetrating into the shocked material of the stellar wind bubble. 
Both vertical and horizontal distributions of the emitted material extend to about 
$40\, \rm pc$ from the center of the explosion, and in thermal X-rays the supernova remnant looks
like a square surrounding the inner optical rectangle. 

%%%%%%%%%%%%%%%%%%%%%%%%%%%%%%%%%%%%%%%%%%%%%%%%%%%%%%%%%%%%%%%%%%%%%%%%%%%%%%%%%%%%%%%%%%%
%%%%%%%%%%%%%%%%%%%%%%%%%%%%%%%%%%%%%%%%%%%%%%%%%%%%%%%%%%%%%%%%%%%%%%%%%%%%%%%%%%%%%%%%%%%
%%%%%%%%%%%%%%%%%%%%%%%%%%%%%%%%%%%%%%%%%%%%%%%%%%%%%%%%%%%%%%%%%%%%%%%%%%%%%%%%%%%%%%%%%%%

\section{Discussion}
\label{sect:discussion}

The caveats and limitations of our study are presented in this section  
and the results are discussed in the context of the rectangular supernova remnant Puppis~A.

\subsection{Limitation of the model}
\label{sect:diss_caveat}

Two main caveats affect our simulation method. First of all, 
regarding our choice in terms of stellar wind boundaries, we neglect the 
magnetization of the progenitor stellar wind, as in the precedent papers of this 
series~\citep{meyer_mnras_493_2020,meyer_mnras_502_2021}, and as discussed in great 
details in~\citet{vanmarle_584_aa_2015}. 
Nevertheless, this is a prerequisite in the careful estimation of cosmic-rays propagation through core-collapse supernova 
remnants~\citep{sushch_apj_926_2022,2022arXiv220303369D}. 
The adopted scenario for the evolution of the progenitor is arbitrarily taken 
to be that of a non-rotating $35\, \rm M_{\odot}$ star with solar 
metallicity~\citep{ekstroem_aa_537_2012}. Hence, this first 
qualitative study for the formation of rectangular middle-age core-collapse supernova 
remnants is not fine-tuned to a particular object 
(see Section~\ref{sect:diss_obsrvation}).

Secondly, we neglect any 
turbulence in the medium 
into which the star blows its wind. In reality, a surrounding \hii region, potentially trapped into the dense layer of the growing stellar 
wind bubble~\citep{weaver_apj_218_1977,vanmarle_revmexaa_22_2004}, develops velocity fluctuations when expanding into an inhomogeneous molecular 
cloud~\citep{medina_mnras_445_2014}. 
Furthermore, the dense environment hosts star formation processes and typically is 
highly magnetized~\citep{hennebelle_arav_20_2012}, influencing the development of 
circumstellar structures~\citep{pardi_mnras_465_2017} and supernova 
remnants~\citep{korpi_aa_350_1999,korpi_apj_514_1999}, which in turn generate and maintain the level of turbulence of the ISM~\citep{seifried_apj_855_2018}. 
A more realistic depiction of the supernova-wind interaction in a magnetized ISM would 
require full 3D magneto-hydrodynamical simulations, better modelling of the circumstellar 
medium around the massive progenitor star~\citep{meyer_mnras_506_2021}, as well as the propagation 
of the supernova blastwave in it~\citep{orlando_aa_645_2021,2022arXiv220201643O}. 
%
%Such task deserves further work exploring the formation scenario of the remnant as well 
%as the conditions of the local ambient medium. 

\subsection{Interpretation and comparison with observations}
\label{sect:diss_obsrvation}

\subsubsection{A new model for rectangular core-collapse remnants}

Our study reveals how core-collapse supernova remnants can reveal a rectangular appearance, by a succession of shock reflections on the walls of the stellar wind cavity, provided their progenitor star blows its pre-supernova material in a medium that is uniformly magnetised~\citep{vanmarle_584_aa_2015}. This mechanism is the adapted version of the interacting-winds model employed for explaining the formation of bipolar and elliptical planetary nebulae, \textcolor{black}{in a sense that the final outcome is attributed to the interaction of a rather spherical outflow with a pre-existing, non-spherical circumstellar structure.}

\textcolor{black}{In that mass regime,}
\textcolor{black}{other models explain the formation of bipolar and multipolar planetary nebulae, characterized by showing point-symmetric features. These models consider a binary system inside \citep{bond1978,livio1979,soker1994}. One of the binary components launches a pair of precessing jets \citep{soker2000} which interact with the AGB wind of the other star. This interaction produces point-symmetric planetary nebulae as observed in the Red Rectangle \citep{cohen2004,velazquez2011}, Hen 3-1475 \citep{riera2014} and IC 4634 nebulae \citep{guerrero2008}. The interacting wind model} considers the collision of an isotropic, fast, and low-density stellar wind (coming from a low-mass star) with a slow and high-density AGB wind, with a high-density equatorial region (shaped by stellar rotation). The wind-wind interaction induces an asymmetric flow and opens the stellar magnetic field lines~\citep{garcia_segura_apj_544_2000}. 
\textcolor{black}{
Analogically, in our model involving a massive progenitor star, the spherical outflow is represented by the supernova ejecta, while the non-spherical circumstellar medium is sculptured by both the red supergiant and Wolf-Rayet stellar 
winds under the action of ISM magnetic fields.
}

We interpret the formation of such rectangular core-collapse supernova remnants as indication of slow or absent motion of the progenitor star.  
If the massive star were a fast-moving object, the cavity of stellar wind would be strongly asymmetric, and so would be the propagation of the shock wave ~\citep{meyer_mnras_450_2015}. 
\\
\\
%{\bf Alex: I suggest the following alternative for Section 4.2.1}

%\textcolor{blue} {Our study reveals how core-collapse supernova remnants can reveal a rectangular appearance, by a succession of shock reflections on the walls of the stellar wind cavity, provided their progenitor star blows its pre-supernova material in a medium that is uniformly magnetised~\citep{vanmarle_584_aa_2015}. This mechanism is the adapted version of the interacting-winds model employed for explaining the formation of bipolar and elliptical planetary nebulae, in a sense that the final outcome is attributed to the interaction of a rather spherical outflow with a pre-excisting, non-spherical circumstellar structure.}

%\textcolor{blue} {The interacting wind model considers the collision of an isotropic, fast, and low-density stellar wind that accompanies the contraction of the stellar core toward the formation of a White Dwarf with the slow, high density and equatorial confined AGB wind~\citep{garcia_segura_apj_544_2000}. Analogically, in our model the spherical outflow is represented by the SN ejecta, while the non-spherical CSM is sculptured by the RSG and WR winds under the action of magnetic fields. }

%\textcolor{blue} {We interpret the formation of such rectangular core-collapse supernova remnants as indication of slow or absent motion of the progenitor star.  
%If the massive star were a fast-moving object, the cavity of stellar wind would be strongly asymmetric, and so would be the propagation of the shock wave ~\citep{meyer_mnras_450_2015}. }

\subsubsection{Comparison with Puppis A}

\begin{figure}
        \centering
        \includegraphics[width=0.45\textwidth]{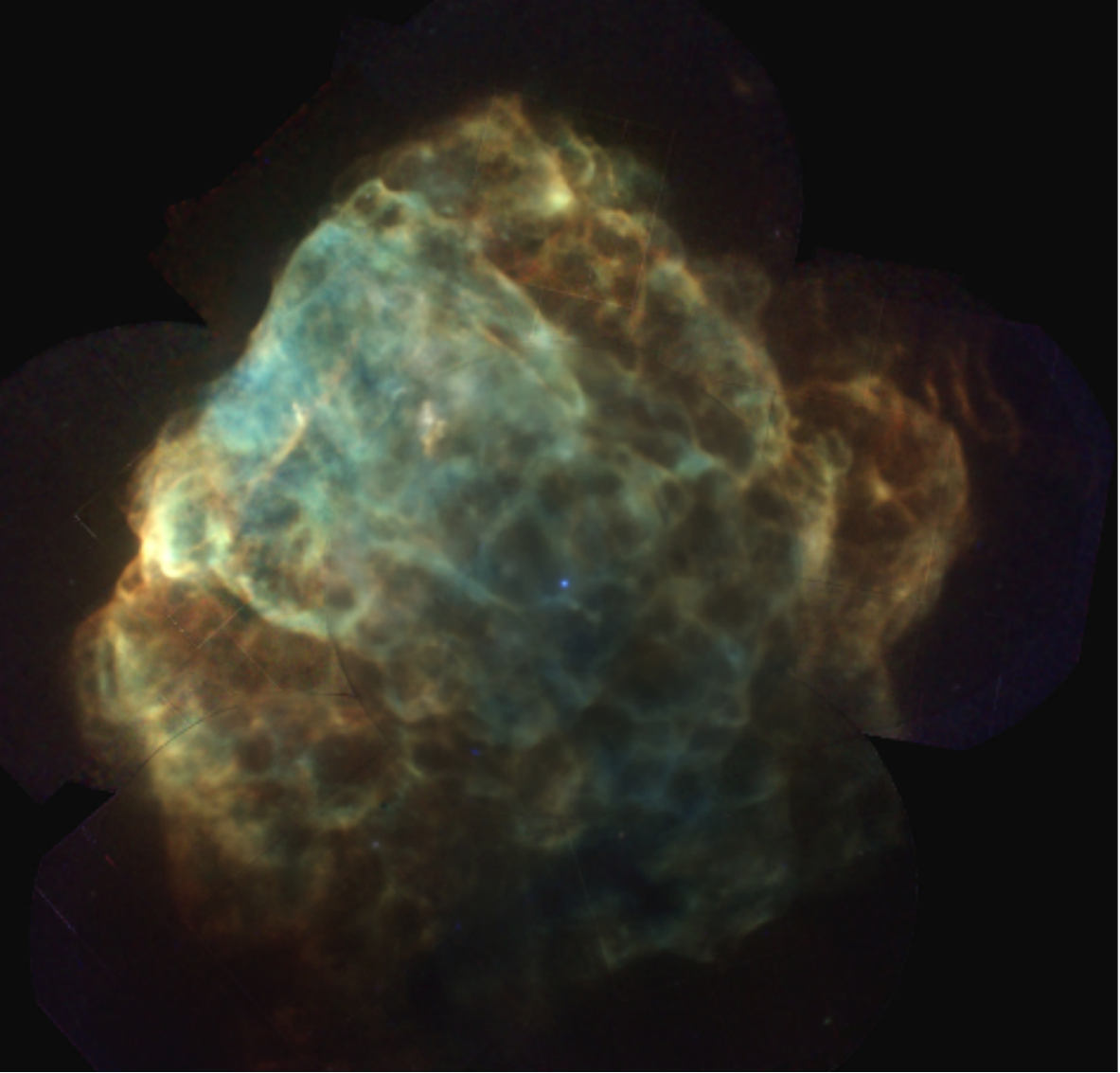}  \\        
        \caption{
        X-ray view of the supernova remnant Puppis~A as a mosaic of XMM-Newton 
        and Chandra data, taken from~\citet{dubner_aa_555_2013}, where the
        red ($0.3-0.7\, \rm keV$), green ($0.7-1.0\, \rm keV$) and blue 
        ($1.0-8.0\, \rm keV$) colors indicate each emission energy band.
        The supernova remnant exhibits a morphology strongly deviating from sphericity, 
        appearing as an overall rectangular shape inside of which multiple  
        square structures are imbricated. 
        }
        \label{fig:puppis_A}  
\end{figure}

Puppis A is the archetype of core-collapse supernova remnant  with a rectangular morphology. 
It displays an unusual angular shape shown in a composite X-ray rendering (see Fig.~\ref{fig:puppis_A}) 
and a rectangular morphology in radio continuum \citep[see Fig.1 in][]{reynoso_mnras_464_2017}, 
with an intensity enhancement toward the Eastern region. 
\textcolor{black}{
The massive nature of Puppis A's progenitor has been constrained thanks to (i) its O and Fe 
enriched interior~\citep{chqrles_mnras_185_1978,katsuda8apj_768_2013} and (ii) the 
runaway pulsar it hosts~\citep{2017ApJ...844...84H,vogt_natas_2018,mayer_apj_899_2020}. 
This supernova remnant has mainly been observed in the 
radio~\citep{reynoso_mnras_464_2017}, 
infrared~\citep{arendt_apj_368_1991,arendt_apj_725_2010,arendt_apj_368_1991}, 
optical~\citep{goudis_aa_62_1978}, 
and X-Ray~\citep{dwek_apj_320_1987,dubner_aa_555_2013} 
band, providing a multi-wavelength picture of the rectangular shape. 
Infrared emission from scattered light on dust trapped into the supernova remnant 
reveals that the supernova blastwave is not expanding into a stratified medium, except 
in the eastern region~\citep{1990ApJ...350..266A}, which is in accordance with our 
model. 
}

\textcolor{black}{
High-energy data provided by the {\it Fermi-LAT} and {\it H.E.S.S.} facilities 
indicate that Puppis A is the site of particle acceleration up to the GeV band only, 
which supports the scenario of the supernova blastwave not yet massively 
interacting with the surrounding dense 
molecular clouds~\citep{hewitt_apj_759_2012,2015A&A...575A..81H,xin_apj_843_2017}. 
This fact favours the rectangular morphology being principally produced by the blastwave 
colliding with circumstellar material, as proposed our model. In particular, a blastwave passing through hot, shocked circumstellar material tends to produce soft cosmic ray spectra \citep{2022A&A...661A.128D}, leading to much weaker emission in the TeV band than at GeV energies, as is observed. 
The sole signs of isolated cloud collision in the bright eastern knot of 
Puppis A~\citep{paron_aa_480_2008}, not included into our simulation, are therefore 
not responsible for its overall shaping. 
}

Our numerical model is consistent with the observations of Puppis~A, as it displays 
an overall squared X-ray shape (Fig.~\ref{fig:plot_IC}d,h) with 
smaller imbricated optical rectangles (Fig.~\ref{fig:plot_IC}c,g). 
This finding permits us to propose that the progenitor of Puppis A was a massive star 
which evolved in a background medium of organised magnetisation in agreement with the 
scenario presented by \citet{reynoso_mnras_477_2018}.

Throughout stellar evolution, the many stellar evolutionary phases released first main-sequence 
material that first shape an elongated cavity in which an evolved red supergiant wind is 
blown~\citep{reynoso_mnras_464_2017,reynoso_mnras_477_2018}, and, depending on the mass 
of the progenitor, it might have further evolved up to the Wolf-Rayet phase.  
\textcolor{blue}{An extra numerical effort is necessary to fine tune} simulations to the case 
of Puppis~A and its shaping.
%, as well as other objects such as bilateral supernova remnants~\citep{gaensler_apj_493_1998}.  

%%%%%%%%%%%%%%%%%%%%%%%%%%%%%%%%%%%%%%%%%%%%%%%%%%%%%%%%%%%%%%%%%%%%%%%%%%%%%%%%%%%%%%%%%%%
%%%%%%%%%%%%%%%%%%%%%%%%%%%%%%%%%%%%%%%%%%%%%%%%%%%%%%%%%%%%%%%%%%%%%%%%%%%%%%%%%%%%%%%%%%%
%%%%%%%%%%%%%%%%%%%%%%%%%%%%%%%%%%%%%%%%%%%%%%%%%%%%%%%%%%%%%%%%%%%%%%%%%%%%%%%%%%%%%%%%%%%

\section{Conclusion}
\label{sect:conclusion}

This study investigates the possibility that core-collapse supernova remnants adopt a rectangular 
emission morphology. We perform 2.5D magneto-hydrodynamical simulations of the circumstellar 
medium of a $35\, \rm M_{\odot}$ single massive star at rest~\citep{ekstroem_aa_537_2012} 
evolving and dying in the warm phase of the Galactic plane of the Milky Way. 
The numerical models are performed with the {\sc pluto} 
code~\citep{mignone_apj_170_2007,migmone_apjs_198_2012,vaidya_apj_865_2018}, 
which has been previously used to study the surroundings of massive 
stars and their associated supernova remnants~\citep{meyer_2014bb}. 

It is known that magnetic field of order of few $\mu G$ is not effective to modify the shape of supernova remnant because the ratio of the magnetic pressure to the thermal pressure is small and magnetic field is not dynamically important in evolution of the remnant. However, it is very effective to determine the structure of the ambient density where this ratio is considerably larger than unity~\citep{vanmarle_584_aa_2015}.
We demonstrate that such circumstellar medium of the progenitor star, profoundly pressure-supported by an organized 
background magnetic field, subsequently constrains the propagation of the supernova shock wave, 
inducing multiple shock reflections along the directions normal to the ISM magnetic fields.  
Eventually, in the middle age of its evolution ($15$$-$$20\, \rm kyr$), the remnant adopts the unusual rectangular morphology due to the isotropically-expanding supernova ejecta interacting with the tubular structure of the progenitor surroundings. 

We calculated intensity maps of radio synchrotron and inverse Compton emission using radiative transfer calculations. At some viewing angles, these maps show a rectangular morphology for the age window between the reflection of the supernova shock wave off the circumstellar medium and the moment it reaches the centre of the remnant. This mechanism is the adapted version of the shaping of bipolar and elliptical planetary nebula by the interaction of two stellar winds \citep[see][]{garcia_segura_apj_544_2000}. Our scenario for rectangular core-collapse supernova remnants requires that its
progenitor was not a runaway star, for which the remnant will shape following the scenario of~\citet{meyer_mnras_502_2021}. 
We suggest that the asymmetric supernova remnant Puppis~A, depicting a 
rectangular morphology, is at least partly shaped by the mechanism described here.

Further investigations are necessary to validate our model for rectangular core-collapse 
supernova remnants, both on the observational and on the numerical side. The former 
to update our knowledge on the abundance of rectangular remnants from massive 
stars, the latter to understand the detailed morphology of 
Puppis~A~\citep{reynoso_mnras_464_2017}.

%%%%%%%%%%%%%%%%%%%%%%%%%%%%%%%%%%%%%%%%%%%%%%%%%%%%%%%%%%%%%%%%%%%%%%%%%%%%%%%%%%%%%%%%%%%
%%%%%%%%%%%%%%%%%%%%%%%%%%%%%%%%%%%%%%%%%%%%%%%%%%%%%%%%%%%%%%%%%%%%%%%%%%%%%%%%%%%%%%%%%%%
%%%%%%%%%%%%%%%%%%%%%%%%%%%%%%%%%%%%%%%%%%%%%%%%%%%%%%%%%%%%%%%%%%%%%%%%%%%%%%%%%%%%%%%%%%%

\section*{Acknowledgements}

\textcolor{black}{
The authors thank the anonymous referee for comments which improved the quality of the paper.
}
DMA Meyer thanks L.~Oskinova for discussion on X-ray observational data. 
The authors acknowledge the North-German Supercomputing Alliance (HLRN) for providing HPC 
resources that have contributed to the research results reported in this paper. 
M.~Petrov acknowledges the Max Planck Computing and Data Facility (MPCDF) for providing data 
storage resources and HPC resources which contributed to test and optimise the {\sc pluto} code. P.F. Velázquez, J.C. Toledo-Roy, and A. Castellanos-Ramírez acknowledge the financial support for PAPIIT-UNAM grants IA103121 and IG100422. A. Castellanos-Ram\'irez acknowledges support from CONACyT postdoctoral fellowship. A. Camps-Fariña acknowledges financial support by the Spanish Ministry of Science and Innovation through the research grant PID2019-107427GB-C31. E.M.~Reynoso is member of the Carrera del Investigador Cient\'\i fico of CONICET, Argentina, and is partially funded by CONICET grant PIP 112-201701-00604CO.

\section*{Data availability}

This research made use of the {\sc pluto} code developed at the University of Torino  
by A.~Mignone (http://plutocode.ph.unito.it/) 
and of the {\sc radmc-3d} code developed at the University of Heidelberg by C.~Dullemond 
(https://www.ita.uni-heidelberg.de/$\sim$dullemond/software/radmc-3d/).
The figures have been produced using the Matplotlib plotting library for the 
Python programming language (https://matplotlib.org/). 
The data underlying this article will be shared on reasonable request to the 
corresponding author.

%%%%%%%%%%%%%%%%%%%%%%%%%%%%%%%%%%%%%%%%%%%%%%%%%%%%%%%%%%%%%%%%%%%%%%%%%%%%%%%%%%%%%%%%%%%
%%%%%%%%%%%%%%%%%%%%%%%%%%%%%%%%%%%%%%%%%%%%%%%%%%%%%%%%%%%%%%%%%%%%%%%%%%%%%%%%%%%%%%%%%%%
%%%%%%%%%%%%%%%%%%%%%%%%%%%%%%%%%%%%%%%%%%%%%%%%%%%%%%%%%%%%%%%%%%%%%%%%%%%%%%%%%%%%%%%%%%%

\bibliographystyle{mnras}
\bibliography{grid} % if your bibtex file is called example.bib

\bsp	% typesetting comment
\label{lastpage}
\end{document}